\documentclass[letterpaper,10pt]{article} 
\usepackage{graphicx} 
\usepackage{amsmath}
\usepackage{siunitx}
\usepackage{comment}
\usepackage[left=1cm,top=1cm,right=1cm,bottom=1cm,includefoot,heightrounded]{geometry}
\usepackage{float}
\usepackage{authblk}

\title{Broadband acousto-optic modulators on Silicon Nitride}
\author[1]{Scott E. Kenning}
\author[1]{Tzu-Han Chang}
\author[1]{Alaina G. Attanasio}
\author[2]{Warren Jin}
\author[2]{Avi Feshali}
\author[1]{Yu Tian}
\author[2]{Mario Paniccia}
\author[1]{Sunil A. Bhave}
\affil[1]{OxideMEMS Lab, Purdue University, West Lafayette, Indiana 47907, USA}
\affil[2]{Anello Photonics, Santa Clara, California 95054, USA}
\begin{document}

\maketitle

\begin{abstract}
    Stress-optic modulators are emerging as a necessary building block of photonic integrated circuits tasked with controlling and manipulating classical and quantum optical systems. While photonic platforms such as lithium niobate and silicon on insulator have well developed modulator ecosystems, silicon nitride so far does not. As silicon nitride has favorable optical properties, such as ultra-low-loss and a large optical transparency window, a rich ecosystem of potential photonic integrated circuits are therefore inhibited. Here we demonstrate a traveling wave optically broadband acousto-optic spiral modulator architecture at a wavelength of $\SI{1550}{\nano\meter}$ using $\SI{90}{\nano\meter}$ thick silicon nitride waveguides and demonstrate their use in an optomechanical sensing system. The spiral weaves the light repeatedly through the acoustic field up to 38 times, factoring in the time evolution of the acoustic field during the light's transit through spirals up to \SI{26}{\centi\meter} in length. These modulators avoid heterogeneous integration, release processes, complicated fabrication procedures, and modifications of the commercial foundry fabricated photonic layer stack by exploiting ultra-low-loss waveguides to enable long phonon-photon interaction lengths required for efficient modulation. The design allows for thick top oxide cladding of $\SI{4}{\micro\meter}$ such that the low loss optical properties of thin silicon nitride can be preserved, ultimately achieving a $V_\pi$ of $\SI{8.98}{\volt}$ at $\SI{704}{\mega\hertz}$ with $\SI{1.13}{\decibel}$ of insertion loss. Our modulators are the first optically broadband high frequency acousto-optic modulators on thin silicon nitride, and the novel architecture is accessible to any low loss photonic platform. We demonstrate an immediate use case for these devices in a high-Q optomechanical sensing system.
\end{abstract}

\section{Introduction}
\begin{figure*}[!htb]
    \centering
    \includegraphics[width=\linewidth]{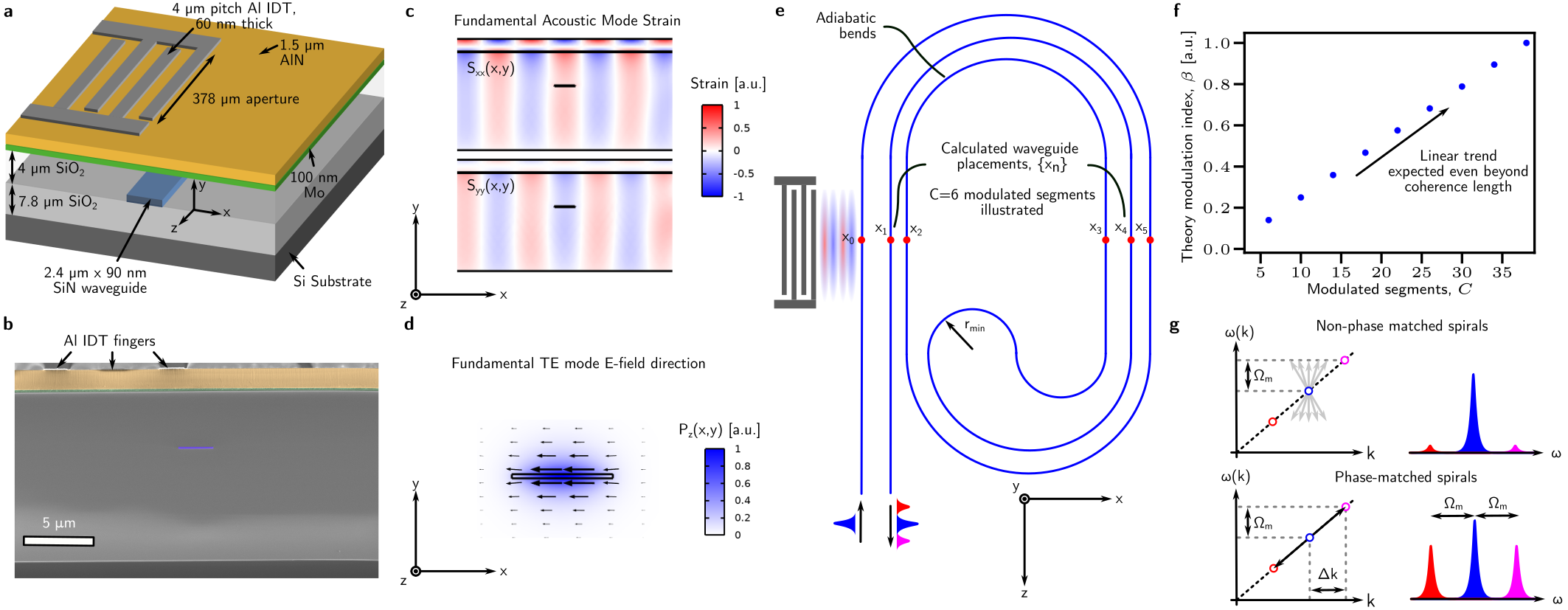}
    \caption{{\bf{Overview of the modulator implementation and design}} | {\bf{a.}} A illustration of the ultra-low-loss foundry-fabricated thin silicon nitride layer stack and subsequently deposited and fabricated piezoelectric transducer layers. The silica cladding above the waveguide is rendered transparent in the illustration so the waveguide is visible. {\bf{b.}} A false colored cross-section scanning electron microscope image of the layer stack illustrated in ({\bf{a}}). {\bf{c.}} Simulation of the relevant strain tensor components, $S_{xx}$ and $S_{yy}$ for the fundamental transducer acoustic mode. {\bf{d.}} An optical mode profile with arrows indicating the electric field is mostly $\hat{x}$-directed. {\bf{e.}} An illustration of the design strategy. The $x$ location of the actively modulated waveguide sections are calculated such that the modulation is coherent in all modulated segments (straight waveguide sections) and design rules such as minimum bending radii are obeyed. {\bf{f.}} The expected behavior of the spiral designs presented in this work. Modulation index is expected to trend linearly with respect to modulated segment count even beyond the coherence length. {\bf{g.}} Dispersion diagram illustrating the effect of group delay compensation in the design of the spiral. By placing waveguides in the spiral according to the proposed method, phase matching can be achieved. In general, most spiral geometries will not satisfy phase matching, and performance will be sub-optimal, limiting modulation indices achievable. }
    \label{fig:fig1}
\end{figure*}
Silicon nitride (SiN) photonics has enabled narrow linewidth lasers \cite{jin_hertz-linewidth_2021, xiang_3d_2023, liu_fully_2024, li_high-coherence_2023, isichenko_sub-hz_2024}, record setting high-Q resonators \cite{puckett_422_2021, liu_ultralow_2022}, and sensing systems \cite{srinivasan_design_2014, tian_high-resolution_2023}. With the further inclusion of piezoelectrics alongside SiN photonic integrated circuits (PICs) \cite{tian_piezoelectric_2024}, fast tunable lasers \cite{lihachev_low-noise_2022, siddharth_piezoelectrically_2024, lihachev_frequency_2024}, soliton microcomb tuning \cite{liu_monolithic_2020}, optical isolators \cite{tian_magnetic-free_2021, cheng_terahertz-bandwidth_2025}, and quantum microwave-to-optical conversion \cite{blesin_bidirectional_2024} have been demonstrated. All of these technologies are building blocks for PIC designers to use for producing mass-manufacturable integrated systems. However, SiN, a material platform noted for its broad optical transparency window \cite{xiang_silicon_2022} reaching the visible spectrum \cite{morin_cmos-foundry-based_2021, corato-zanarella_absorption_2024}, low optical losses \cite{puckett_422_2021}, and CMOS compatibility \cite{jin_hertz-linewidth_2021} suffers from a lack of a generally applicable optically broadband modulator architecture. Modulators have been demonstrated through heterogeneous integration of electro-optic materials such as lithium niobate with SiN \cite{ghosh_acousto-optic_2021, churaev_heterogeneously_2023, badri_compact_2025, rahman_high-performance_2025}, but requires substantial modifications to the cladding layer to bring electro-optic material within proximity to the SiN waveguide core. 
When broadband modulation frequencies are not necessary, acousto-optic modulators are an attractive choice. Laser locking schemes, such as the Pound-Drever-Hall technique \cite{drever_laser_1983}, require optical modulators to generate the error signal used by the control system. PICs have been proposed \cite{isichenko_photonic_2023, blumenthal_enabling_2024} that utilize integrated modulators and lasers to interface with atomic transitions, serving as a necessary next step to developing PICs for light-atom interactions. As will be later demonstrated in this work, they also lend use towards integration of optomechanical accelerometers using high-Q ring resonators. 

Currently, the few existing acousto-optic modulators on SiN rely on optical resonant enhancement of the stress optic interaction \cite{tian_hybrid_2020, wang_silicon_2022}, limiting the optical bandwidth to that of the ring resonator. Recent work on SiN seeks to address these shortcomings by embedding piezoelectric material and an air gap below the optical waveguide to enhance optomechanical coupling and produce power efficient optically broadband modulators \cite{freedman_gigahertz-frequency_2025}. Looking beyond SiN, Si waveguide platforms typically feature released and air clad waveguides \cite{zhou_nonreciprocal_2024, cheng_terahertz-bandwidth_2025}. However, these methods cannot be translated to all SiN platforms, particularly ``thin" SiN with waveguide heights below $\SI{100}{\nano\meter}$ responsible for record low optical losses. Silica cladding for thin SiN is on the order of several microns to drive down losses, and including piezoelectric materials in close proximity the waveguide would spoil the optical loss. 

To address the gaps in SiN modulator technology, we develop a novel and counterintuitive design method for optically broadband phase modulators by exploiting the low optical losses of $\SI{90}{\nano\meter}$ thick waveguides. This technique is general and can be applied to both thick and thin SiN waveguides. The photonic waveguides are fabricated in a commercial foundry and no modifications to the waveguide and cladding layers are made (such as release processes), leading to 100\% yield. Using $\SI{90}{\nano\meter}$ SiN waveguides, our modulators operate up to $\SI{704}{\mega\hertz}$, representing a $35\times$ increase in operation frequency over other work with similar waveguide heights \cite{wang_silicon_2022} while achieving a $V_{\pi}$ of $\SI{8.98}{\volt}$ with $\SI{1.13}{\decibel}$ of insertion loss across a broad optical bandwidth. Additionally, the method places minimal constraints on photonic layer thicknesses to avoid the modulator dictating the layer stack in future PICs based on this technology. Trends expected from theory are experimentally measured to confirm the phase-matching physics of the modulators. Finally, we demonstrate an immediate application of the technology to an optomechanical acceleration sensing system based on high-Q ring resonators fabricated at the same commercial foundry.

\section{Methods}

Broadband acousto-optic modulation on a $\SI{90}{\nano\meter}$ SiN waveguide platform is challenging because of the low-confinement waveguides. Techniques involving moving boundary effects in high-confinement Si waveguides originally used to enhance Brillouin interactions \cite{rakich_tailoring_2010, rakich_giant_2012, shin_tailorable_2013} therefore do not translate to thin SiN, relegating acousto-optic interactions to being dependent on the photoelastic effect. Previously, modulators have been produced on this platform using optical ring resonators to yield low power dual sideband modulators \cite{wang_silicon_2022}. However, such devices are not optically broadband and limited in modulation frequency. The challenge with thin SiN waveguides arises due to a large fraction of the optical mode residing in the low index silica cladding. As specified by equation \ref{eq:delta_n} \cite{kittlaus_-chip_2017}, the refractive index shift ($\Delta n_{ij}$) due to strain ($S_{kl}$) is proportional to the photoelastic tensor $p_{ijkl}$ times the material refractive index cubed $n_0^3$.
\begin{equation}\label{eq:delta_n}
    \Delta n_{ij} = -\frac{1}{2}n_0^3 p_{ijkl}S_{kl}
\end{equation}
When thin SiN waveguides are compared to high-index and high-confinement platforms such as silicon waveguides \cite{cheng_terahertz-bandwidth_2025, kittlaus_electrically_2021}, the $n_0^3$ term alone accounts for a $14\times$ difference between the two materials at a wavelength of $\SI{1550}{\nano\meter}$, although the silica tensor does help compensate for this deficiency as its values are larger than that of Si \cite{rakich_tailoring_2010}. In addition to this, the thick top cladding used to ensure the mode does not leak above the chip surface makes perturbative interactions with the light a further challenge. As a result, our modulators require significant optimization of the photonics and transducers. Other SiN work embeds aluminum nitride (AlN) and air gaps within $\SI{710}{\nano\meter}$ of $\SI{300}{\nano\meter}$ thick SiN waveguides \cite{freedman_gigahertz-frequency_2025}, but the low confinement of our $\SI{90}{\nano\meter}$ waveguides preclude modification of the photonic cladding layer without severely harming the ultra-low-loss (ULL) properties. 

The photonics is fabricated at a commercial foundry on $\SI{200}{\milli\meter}$ wafers. They are cored down to $\SI{100}{\milli\meter}$ and a commercial vendor deposits $\SI{1.5}{\micro\meter}$ AlN with a $\SI{100}{\nano\meter}$ Mo seed layer. Aluminum interdigited transducers (IDTs) are fabricated chip-scale in the university cleanroom with a single-mask liftoff of a $\SI{60}{\nano\meter}$ aluminum layer, resulting in a layer stack illustrated in Figure \ref{fig:fig1}a with a false-colored SEM in Figure \ref{fig:fig1}b. IDTs have previously been used for modulating waveguides on a variety of optical platforms, such as silicon-on-insulator \cite{kittlaus_-chip_2017, erdil_wideband_2024}, aluminum nitride \cite{tadesse_sub-optical_2014, vainsencher_bi-directional_2016}, and lithium niobate \cite{cai_acousto-optical_2019, jiang_efficient_2020, sarabalis_acousto-optic_2021, sohn_electrically_2021}. The initial transducers are single layer to decrease fabrication variation of transducers across the chip, allowing the modulation properties of multiple modulators to be probed without introducing effects due to transducer non-uniformity. Later, two metal layers of differing thicknesses are used to further optimize electromechanical impedance matching. 

The modulators comprise of long spirals of waveguide (up to $\SI{26}{\centi\meter}$ presented in this work) fabricated on a foundry SiN photonics process to increase the acousto-optic interaction length in a compact area. Spirals are seeing increased use in PICs to enhance acousto-optic interaction, most notably in Brillouin spirals  \cite{morrison_compact_2017, garrett_integrated_2023, ye_brillouin_2024, zhou_nonreciprocal_2024, klaver_surface_2024, neijts_-chip_2024}, where acoustic waves (including surface acoustic waves) co-propagate with the light. Our spirals expand on this paradigm by utilizing acoustic waves traveling perpendicular to the modulated segments of waveguide to formulate a scalable and novel approach to intraband multi-pass acousto-optic modulation. 

With long interaction lengths above the coherence length present in the spiral, phase matching must be taken into account. For a radial modulation frequency $\Omega_m$, group index $n_g \approx 1.57$, and $c_0$ light speed in vacuum, a phase mismatch of
\begin{equation}
\Delta k = \frac{\Omega_m n_g}{c_0}
\end{equation}
is accumulated per length of interaction in a single modulated waveguide segment. The modulation efficiency (away from the carrier suppression regime) is proportional to
\begin{equation}
\eta^2 \propto L^2 \text{sinc}^2\left({\frac{\Delta k L}{2}}\right) = L^2 \text{sinc}^2\left({\frac{L}{L_\text{coh}}}\right)
\end{equation}
where
\begin{equation}\label{eq:coherence_length}
L_{\text{coh}} \equiv \frac{2}{\Delta k} = \frac{2 c_0}{\Omega_m n_g} \approx \SI{13}{\centi\meter}.
\end{equation} 
is the coherence length \cite{boyd_nonlinear_2020}. Modulation efficiency, $\eta^2$, can be related to modulation index, $\beta$, through the Bessel functions.
\begin{equation}
    \eta^2 \propto \left(J_1(\beta)\right)^2
\end{equation}
The instantaneous phase of the light on the output of our modulators is
\begin{equation}
\phi(t) = \beta \sin(\Omega_m t).
\end{equation}

The device architecture consists of dozens of $\SI{1.6}{\milli\meter}$ modulated waveguide segments each far below the coherence length. To allow for the total spiral length to be above the coherence length, the modulated segments are separated by delay lengths formed from adiabatic sine-cosine waveguide bends. The modulated waveguide segments are the straight waveguide sections in the spiral that intersect the acoustic wave, illustrated in Figure \ref{fig:fig1}e. Although only $C=6$ modulated segments are illustrated, this method generalizes to an arbitrary number of modulated segments (up to $C=38$ demonstrated in this work). With proper lithographic phase control between each modulated segment \cite{kittlaus_electrically_2021}, the modulation index achievable should scale linearly with modulated segment count, as shown in Figure \ref{fig:fig1}f. Viewed as a dispersion diagram in Figure \ref{fig:fig1}g, failure to account for phase matching in the spirals results in weak modulation and limitations on the waveguide lengths of the spiral. However, with phase matching the modulation coherently scatters light to another location on the dispersion curve. The modulated segments of waveguides are placed at a calculated position in the acoustic field such that the adiabatic waveguide bends themselves serve as delay lines. A tractable method to design the structure, factoring in geometric constraints such as minimum bending radii, is discussed in detail in the SI. Effectively, our devices bridge the five-order of magnitude gap between the acoustic wave velocity and the light through use of delay lines and an extremely optimized design. A group of light propagating through the spiral intersects the same phase of the acoustic wave in all passes through the acoustic field, leading to coherent modulation. The modulated segment count for our spirals present on the chip was selected such that the device behavior can be experimentally validated to agree with theoretical trends expected as the spiral surpasses the coherence length.

The devices initially demonstrate modulation using the fundamental acoustic mode to place minimal constraints on the layer stack and increases the generality of the approach. The top cladding thickness measured to be $\SI{3.98}{\micro\meter}$ was selected to keep optical losses incurred by the molybdenum layer below $\SI{5}{\decibel\per\meter}$, with measured values obtained from fits to an optical ring resonator indicating less than $\SI{4.31}{\decibel\per\meter}$ across a wide optical bandwidth (experimental loss measurements in SI). This indicates the largest spiral in this work has an estimated \SI{1.13}{\decibel} of insertion loss, which is dominated by metal-induced propagation loss. The silica photoelastic tensor, $p_{ijkl}$ in equation \ref{eq:delta_n}, indicates that the acousto-optic interaction is dominated by the $S_{yy}$ component of the strain tensor, simulated in Figure $\ref{fig:fig1}c$, and the TE mode of the waveguide, simulated in Figure $\ref{fig:fig1}d$. As the acousto-optic interaction on thin SiN is extremely weak compared to other platforms, waveguides must intersect the acoustic wave many times to achieve useful modulation indices. Carefully designed group-delay compensated spirals, shown in Figure \ref{fig:fig1}e, can achieve this.

A traveling wave design was counterintuitively selected over a mechanically resonant design, such as bulk acoustic wave (BAW) resonances \cite{tian_magnetic-free_2021, tian_hybrid_2020}, because of the constraints placed by phase-matching. BAW resonances in the silica appear in the few-GHz regime, and the coherence length of the interaction would therefore be on the order a few $\SI{}{\centi\meter}$. Due to the weakness of the stress-optic effect on this platform, waveguide lengths would necessarily be larger than the coherence length and phase matching would limit device performance. Traveling acoustic waves enable control of the acoustic phase seen by each actively modulated segment of waveguide, preventing these devices from being limited by coherence length. In principle, designers can use this method to arbitrarily reduce $V_{\pi}$ with the trade off of insertion loss and area.

\section{Results}

\begin{figure*}[!htb]
    \centering
    \includegraphics[width=\linewidth]{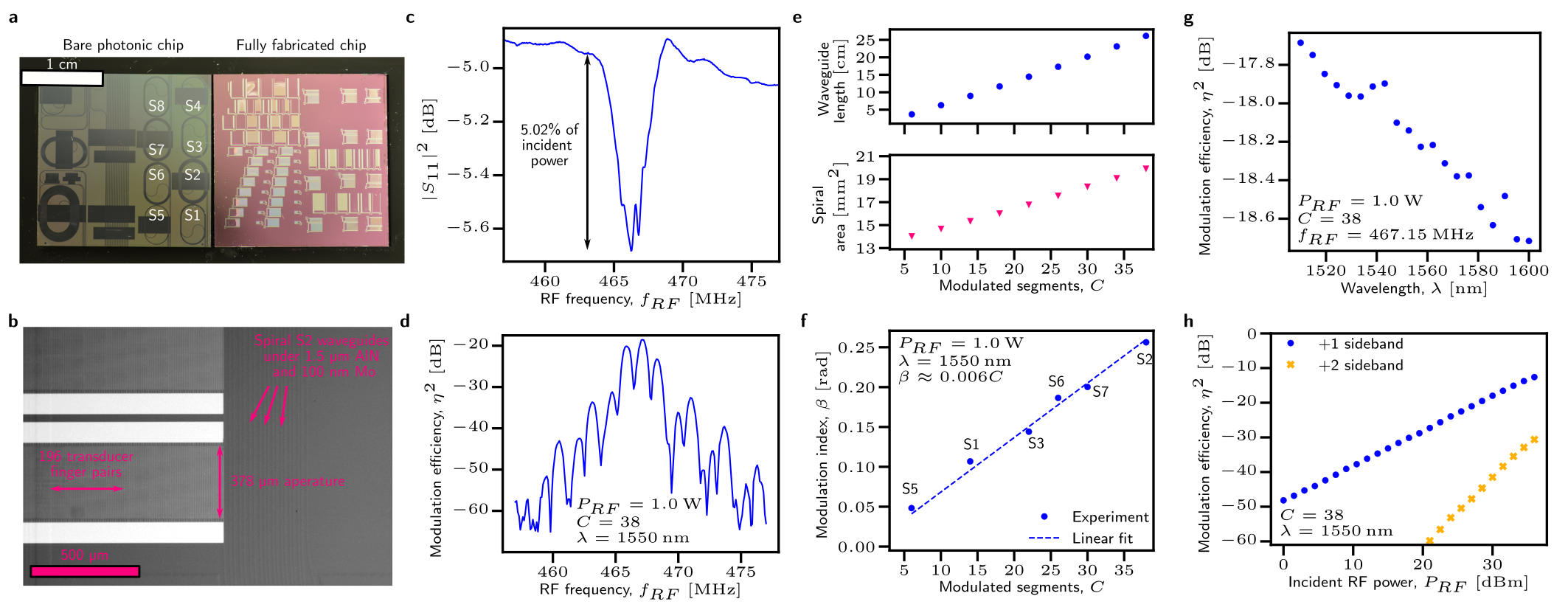}
    \caption{{\bf{Validation of the modulator design |}} {\bf{a.}} A chip without AlN deposited so the photonics are visible (left) and a completely fabricated chip with AlN and transducers (right). {\bf{b.}} An optical microscope image of a transducer next to a spiral. There are 19 waveguides visible on the transducer side of the spiral, corresponding to $C=38$ modulated segments when counting both sides. The image color has been modified such that the waveguides are more easily distinguishable. {\bf{c.}} $|S_{11}|^2$ of the transducer revealing electro-acoustic coupling. {\bf{d.}} Modulation efficiency as a function of transducer RF drive frequency. The fringing pattern validates that both the near and far side of the spiral relative to the transducer both contribute to the modulation. {\bf{e.}} Scaling of spiral geometric parameters as a function of modulated segment count. Note that the spiral area is defined as the area of the smallest rectangle it can fit in. {\bf{f.}} Multiple spirals of varying pass count, $C$, are measured with identical transducer designs to ensure behavior follows the expected linear trend presented in Figure \ref{fig:fig1}c. {\bf{g.}} Broadband modulation efficiency as a function of wavelength, showing $\SI{1.1}{\decibel}$ variation across a $\SI{90}{\nano\meter}$ range. {\bf{h.}} Scaling of the modulation efficiency, for the first and second sidebands with respect to applied incident RF power, following the expected trend for acousto-optic modulation. The negative sidebands track identically to their positive counterparts.}
    \label{fig:fig2}
\end{figure*}

{\bf{Validation of the spiral design:}} A fully fabricated test chip is shown in Figure \ref{fig:fig2}a next to a bare photonic chip where the actual spiral structures are visible. The number of modulated segments in each spiral, ranging from 6 to 38, was selected so that the expected trend in Figure \ref{fig:fig1}f could be experimentally validated beyond the coherence length of $\SI{13}{\centi\meter}$ for the fundamental transducer mode. Figure \ref{fig:fig2}b shows a optical microscope image of the transducers used to illustrate relevant geometries. The transducers' fundamental mode arises at $\SI{467}{\mega\hertz}$ as shown by the $|S_{11}|^2$ in Figure \ref{fig:fig2}c indicating the conversion of electrical energy into an acoustic wave. The modulation efficiency, defined as the ratio in optical power in the carrier and first sideband \cite{kittlaus_electrically_2021}, was measured interferometrically and peaks at the expected fundamental acoustic mode frequency of $\SI{467}{\mega\hertz}$ in Figure \ref{fig:fig2}d. The smallest fringing pattern ($\SI{1.2}{\mega\hertz}$) arises due to coherent modulation of the near and far side of the spiral (separated by $\SI{2.8}{\milli\meter}$) relative to the transducer. Acoustic losses are experimentally measured to be $\SI{0.54}{\decibel/\milli\meter}$ for the fundamental transducer mode (see SI), so it is advantageous to ensure both sides of the spiral are phase matched because significant acoustic power reaches the far side. To validate the design methodology, the modulation index must increase linearly with modulated segment count above the coherence length. Spiral geometric information, presented in Figure $\ref{fig:fig2}$e, indicates that the coherence length of $\SI{13}{\centi\meter}$ corresponds to an modulated segment count of approximately 18. Indeed, the linear trend does continue with the experimentally collected data in Figure $\ref{fig:fig2}$f. 

The modulators are optically broadband (Figure \ref{fig:fig2}g) with modulation efficiency varying by $\SI{1.1}{\decibel}$ across the tuning range of our laser. The group velocity of the fundamental TE mode of the waveguide varies minimally over the $\SI{1500}{\nano\meter}$ to $\SI{1600}{\nano\meter}$, leading to this behavior. The acousto-optic nature of the interaction can be seen by examining the slope of modulation efficiency for the 1st and 2nd order sidebands in response to an RF power sweep, shown in Figure \ref{fig:fig2}h. The first and second sidebands scale with $1\times$ and $2\times$ the applied RF power in $\SI{}{\decibel\text{m}}$.

\begin{figure*}[!htb]
    \centering
    \includegraphics[width=\linewidth]{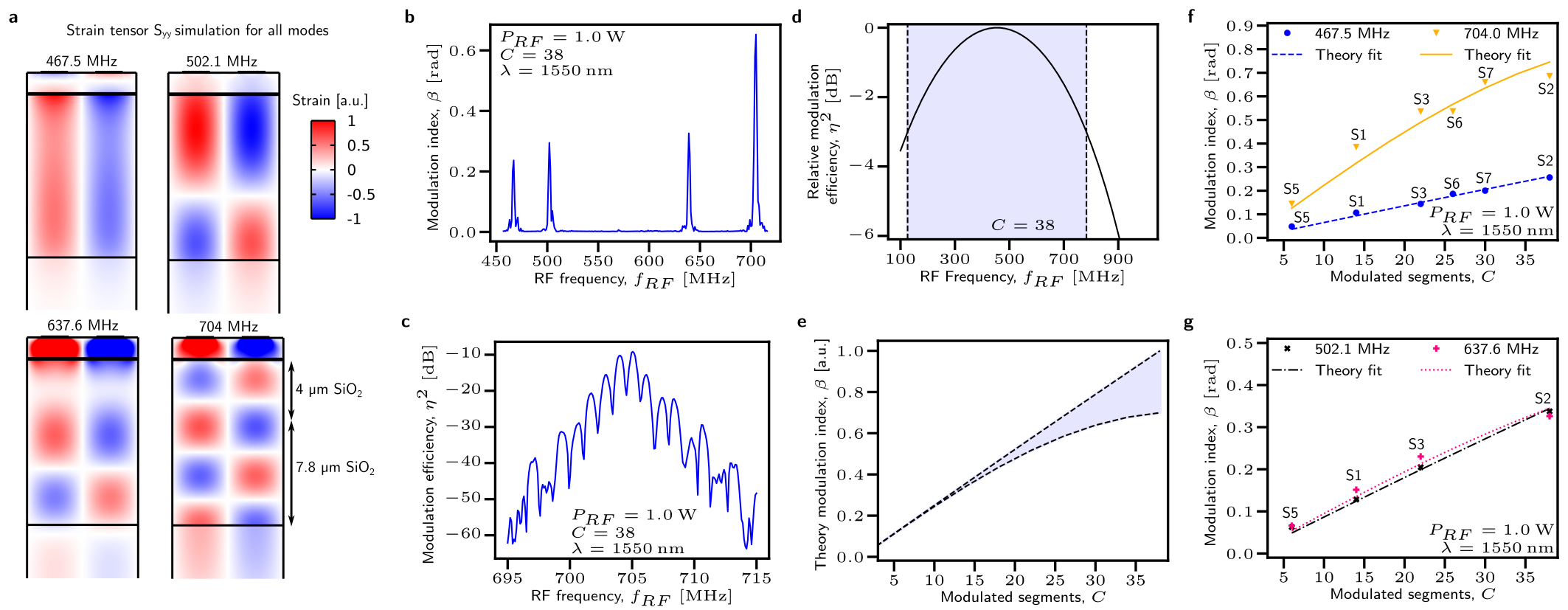}
    \caption{{\bf{Higher-order acoustic modes |}} {\bf{a.}} $S_{yy}$ strain tensor components for three other acoustic modes that can be excited by the transducer. Note that the waveguide location is marked, but in actual devices the waveguides do not reside under the transducers. {\bf{b.}} A broadband RF drive frequency sweep revealing higher order acoustic modes coupling to the optical mode. {\bf{c.}} Modulation efficiency for the $\SI{704}{\mega\hertz}$ mode. It is the strongest performing acoustic mode due to the favorable overlap of the $S_{yy}$ strain tensor component with the waveguide shown in ({\bf{a}}). However, it requires tight layer thickness control. {\bf{d.}} Theoretical calculations illustrating relative degradation of modulation efficiency as acoustic mode frequency differs from the spiral's designed frequency. The shaded blue region shows the $\SI{3}{\decibel}$ range, indicating all acoustic modes in {({\bf{a}})} and {({\bf{b}})} are still expected to nearly satisfy phase matching considerations. {\bf{e.}} Theoretical calculation showing the range of possible modulation index trends expected across the $\SI{3}{\decibel}$ shaded region in ({\bf{d}}) as a function of modulated segments. The $\SI{704}{\mega\hertz}$ mode in ({\bf{f}}) exhibits behavior similar to the lower dashed black line. {\bf{f.}} Scaling of the modulation index for the $\SI{704}{\mega\hertz}$ contrasted against the fundamental mode. The agreement of the theory fits to the data validates the behavior claimed in ({\bf{d}}). {\bf{g.}} Modulation index scaling of the $\SI{502}{\mega\hertz}$ and $\SI{639}{\mega\hertz}$ modes. The modulation index scales nearly linearly, as expected from theory.}
    \label{fig:fig3}
\end{figure*}

{\bf{Examination of higher-order acoustic modes}:} The transducers couple to three more acoustic modes whose $S_{yy}$ strain tensor is simulated in Figure $\ref{fig:fig3}$a. Higher order acoustic modes may be beneficial in photonic integrated circuit design if the layer stack can be specified by the modulator design. For these devices, the modulation indices achievable for each mode is experimentally measured in Figure $\ref{fig:fig3}$b. A high frequency mode present at $\SI{704}{\mega\hertz}$ achieves the strongest coupling, yet requires tight layer thickness control (within $\SI{1.5}{\micro\meter}$) as shown by the corresponding plot in Figure $\ref{fig:fig3}a$, lest the waveguide fall into an anti-node. The modulation efficiency of this mode peaks near $\SI{-10}{\decibel}$ (Figure \ref{fig:fig3}c). 

The higher order modes provide a mechanism to probe the degradation of phase matching as a function of RF drive frequency. As the drive frequency deviates from the spiral's designed $\SI{454}{\mega\hertz}$ ideal operation frequency, a penalty in modulation efficiency occurs due to a phase mismatch. This is theoretically shown in Figure \ref{fig:fig3}d. The largest device presented in this work, with 38 modulated segments, has a phase-matched $\SI{3}{\decibel}$ bandwidth of several hundred $\SI{}{\mega\hertz}$, indicating all higher order modes can be viewed as phase-matched given this $\SI{3}{\decibel}$ metric. Note that this does not account for different electromechanical and acousto-optic coupling for each mode. Acoustic modes within this regime will demonstrate nearly linear scaling of the modulation index with modulated segment count. The black dashed lines in Figure $\ref{fig:fig3}e$ show the upper and lower bounds of curve shapes expected by modes in this range. Figure \ref{fig:fig3}e is experimentally demonstrated with Figures \ref{fig:fig3}f and \ref{fig:fig3}g. The theoretically calculated curve for modulation index scaling is fit to the collected data for the three additional modes. All trends are nearly linear, but the $\SI{704}{\mega\hertz}$ mode exhibits the most deviation from the linear trend expected for perfectly phase-matched acoustic modes. Despite the coherence length of $\SI{8.6}{\centi\meter}$ for the $\SI{704}{\mega\hertz}$ mode being even less than the fundamental acoustic mode, roll-off due to phase mismatch in modulation index is minimal. 

\begin{figure*}[!htb]
    \centering
    \includegraphics[width=\linewidth]{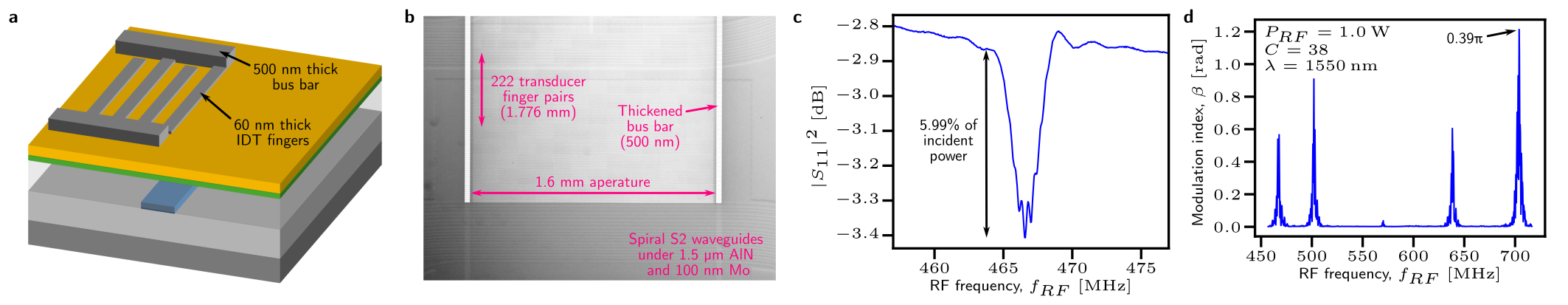}
    \caption{{\bf{Optimization of acousto-optic impedance match |}} {\bf{a.}} An illustration of transducers with thickened bus bars. The layer stack is otherwise identical Fig. \ref{fig:fig1}a. {\bf{b.}} An optical microscope image of the new transducer design with relevant geometric parameters denoted. {\bf{c.}} The $|S_{11}|^2$ of these transducers. This subfigure can be directly compared to Figure \ref{fig:fig2}c to see a greater fraction of incident power is dissipated into the acoustic wave. {\bf{d.}} Modulation index for all acoustic modes present, showing a substantial performance increase when compared to Figure \ref{fig:fig2}b.}
    \label{fig:fig4}
\end{figure*}

{\bf{Optimization of electromechanical impedance match}:} With the spirals validated to agree with theory, the modulation indices may further be increased through transducer design. As speculated by literature, \cite{kittlaus_electrically_2021}, $\SI{}{mm}$-aperture transducers may hold promise in improving acousto-optic modulation. The spirals presented in this work are uniquely positioned to examine this experimentally as the modulated segments in each spiral can accommodate transducers that have a $\SI{1.6}{\milli\meter}$ aperture. The key to large transducers is to thicken the bus bars, shown in Figure \ref{fig:fig4}a. The $\SI{500}{\nano\meter}$ thickness is selected as it is the upper limit of our deposition tool. This drives down parasitic resistance and capacitance simultaneously, allowing finger count and aperture width to be scaled such that the electromechanical coupling component of the impedance is brought closer to $\SI{50}{\ohm}$. A fully fabricated transducer with relevant geometric parameters next to a spiral is shown in Figure \ref{fig:fig4}b. The geometric parameters were arrived at through optimization of an electrical model to improve electromechanical impedance matching for the two lower frequency modes.

The fraction of incident electrical power launched as an acoustic wave for the fundamental mode can be estimated by examining the $|S_{11}|^2$ measurements of the dual metal thickness transducers (Figure $\ref{fig:fig4}$c) and the single metal thickness transducers (Figure $\ref{fig:fig3}$c). A marginal increase of $19.1\%$ is observed. More importantly, this power is distributed over a longer length of waveguide by a factor of $\SI{1.6}{\milli\meter}/ \SI{378}{\micro\meter} = 4.23$. As such, we expect the modulation index achievable for the fundamental acoustic mode to increase by a factor of $\sqrt{4.23\times 1.191} = 2.25$. Measured optically, modulation index increases from by a factor of $2.375$ for the fundamental transducer mode, indicating good agreement between electrical and optical observations. Large transducers are therefore clearly beneficial for acousto-optic modulation. The $\SI{704}{\mega\hertz}$ mode nearly achieves a $\beta=\pi/2~\SI{}{\radian}$ phase shift with $\SI{1}{\watt}$ of RF power, making these devices candidates for the optically broadband modulators used for driving optical gyroscopes \cite{srinivasan_design_2014, tran_integrated_2017}.

\begin{figure*}[!htb]
    \centering
    \includegraphics[width=\linewidth]{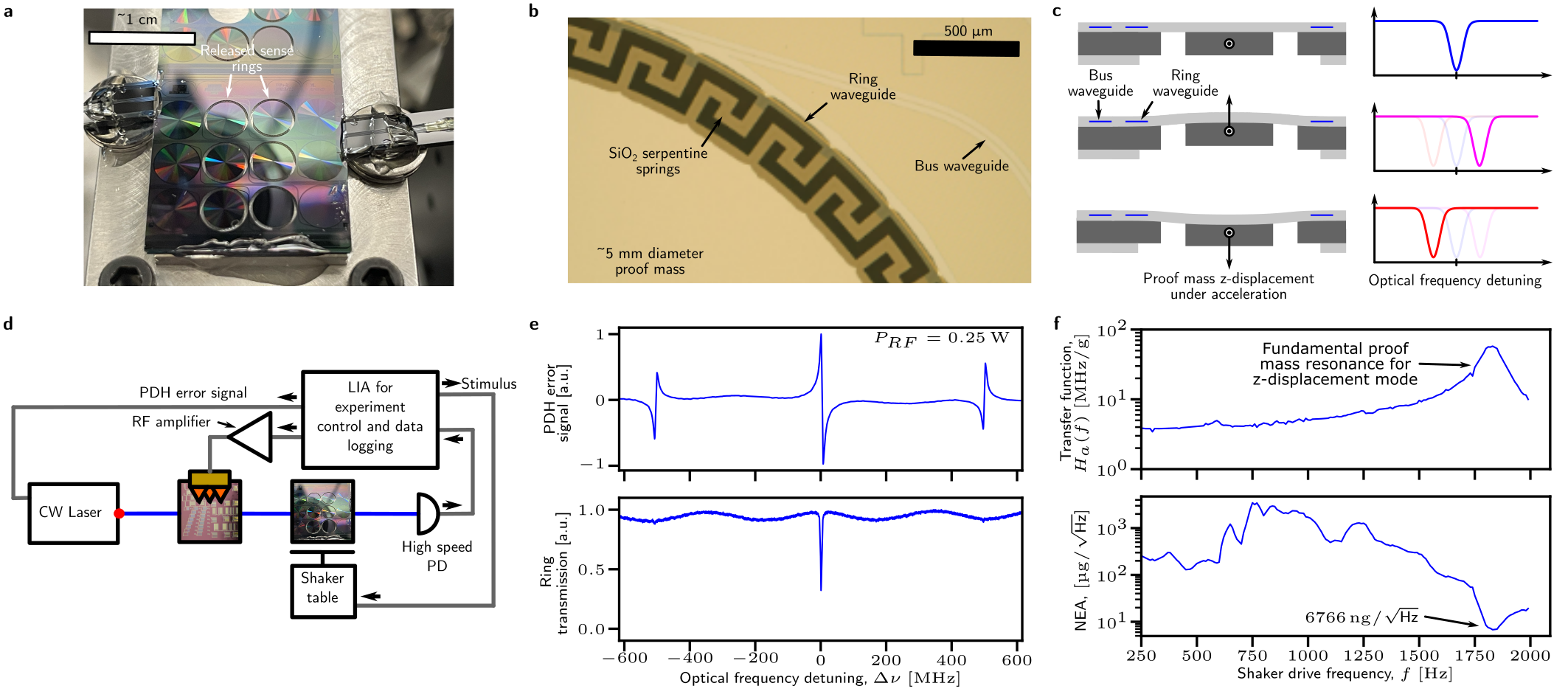}
    \caption{{\bf{Application of modulators to optomechanical sensing system |}} {\bf{a.}} A picture of an optomechanical accelerometer. Four optical ring resonators are accessible via packaging, with two being released and sensitive to acceleration. The chip was produced with an identical photonic layer stack to Figure \ref{fig:fig1}a (however, no piezoelectric was deposited). {\bf{b.}} An optical microscope image of the SiO$_2$ serpentine spring structure that allows the ring resonator to be sensitive to acceleration. {\bf{c.}} A diagram explaining device operation. As the proof mass displaces, the ring resonator is perturbed, shifting the optical resonance frequency. {\bf{d.}} The experiment used to characterize the optomechanical accelerometer. {\bf{e.}} Using $\SI{0.25}{\watt}$ of RF power, a PDH error signal can be generated using the modulators. {\bf{f.}} The transfer function and noise floor is measured. The fundamental resonance of the released mechanical structure is near $\SI{1.8}{\kilo\hertz}$.}
    \label{fig:fig5}
\end{figure*}

{\bf{Use of modulators in a optomechanical sensing system:}} The modulation indices reported in this work are immediately useful. The modulators are used as part of a system to perform Pound-Drever-Hall (PDH) locking \cite{drever_laser_1983} to an optical ring resonator sensitive to acceleration. On a separate chip produced at the same foundry with an identical photonic layer stack, acceleration-sensitive optical ring resonators are fabricated that feature a released proof mass structure on the interior, as shown in Figure \ref{fig:fig5}a. Serpentine silica springs connect the proof mass to a location in proximity to the ring resonator waveguide (Figure \ref{fig:fig5}b). The operation principle of the optical accelerometer is summarized in Figure \ref{fig:fig5}c, where the ring resonator frequency shifts in response to an applied acceleration. The experimental setup uses a Zurich UHFLI lock-in amplifier (LIA) to generate the drive signal for the modulators and demodulate the beat signal. The $\SI{502}{\mega\hertz}$ acoustic mode is used on the chip with dual metal layers. An optical resonance is shown in Figure \ref{fig:fig5}e alongside the corresponding PDH error signal. As the resonance moves in response to acceleration, a nonzero error signal results which is measured by the same LIA to determine the transfer function of the accelerometer and noise equivalent acceleration (Figure \ref{fig:fig5}f). 

Optomechanical accelerometers have been demonstrated on various platforms previously \cite{krause_high-resolution_2012, hutchison_z-axis_2012, li_characterization_2018, zhou_broadband_2021, tian_high-resolution_2023}, but readout schemes rely on side-of-fringe locking schemes and involve a number of components off chip such as photodetectors. PDH locking, enabled by our modulators, rejects intensity noise to first order. Noting that photodetectors and lasers are possible with thin SiN platforms \cite{xiang_3d_2023}, the addition of a simple modulator architecture shown in this work enables fully integrated thin SiN optomechanical accelerometers. 

\section{Discussion}
We have demonstrated, to the best of our knowledge, the first optically broadband acousto-optic modulators using thin SiN waveguides, achieving a $V_\pi$ of $\SI{8.98}{V}$ at $\SI{704}{\mega\hertz}$ with $\SI{1.13}{\decibel}$ of insertion loss. To overcome the weak acousto-optic coupling on this platform, we exploit the low optical losses to compactly coil $\SI{26}{\centi\meter}$ of waveguide into a spiral that can be coherently modulated by an acoustic wave propagating across it. The technique to phase-match long spiral waveguides to traveling acoustic waves is completely novel.  Most importantly, the operation of the devices does not rely on modification of a relatively simple foundry fabricated photonic layer stack through release processes or heterogeneous integration - immediately being broadly applicable to a wide variety of low loss photonic processes. Furthermore, this technology can be trivially extended to reduce RF power requirements and achieve higher modulation indices by increasing modulated segment count. Additionally, calculations indicate that larger transducers with apertures greater than the $\SI{1.6}{\milli\meter}$ demonstrated in this work are possible, indicating that increasing the modulated segment length is the most favorable way to reduce $V_\pi$. Insertion loss can be driven down further through removal of the AlN above the waveguides and recovering the foundry loss \cite{guo_investigation_2024} by using alternative transducer designs such as \cite{plessky_3rd_2013}. This is uniquely enabled by our choice of a traveling wave architecture because the transducer does not need to be above the waveguides. Throughout this work, we have emphasized the generality of this approach to minimize the influence these modulators have in specifying the layer stack. PIC designers can additionally expand on this approach by pursuing more aggressive mode engineering strategies as allowed by other PIC components in their system and fabrication capabilities. A immediate use case for these modulators is demonstrated in prototype system for acceleration sensing with high-Q ring resonators fabricated on the same foundry process, paving the way for immediate use of these devices in a integrated optomechanical accelerometer.

\section{Acknowledgment}
Research was sponsored by the Army Research Laboratory with SEMI-PNT and was accomplished under Cooperative Agreement Number 911NF-22-2-0050. The views and conclusions contained in this document are those of the authors and should not be interpreted as representing the official policies, either expressed or implied, of the Army Research Laboratory or the U.S. Government or SEMI. The U.S. Government is authorized to reproduce and distribute reprints for Government purposes notwithstanding any copyright notation herein. The research at Purdue University is also supported by the U.S. DOE Office of Science, High Energy Physics, QuantISED program (FWP ERKAP63) and the U.S. DOE Office of Science, Quantum Science Center. This material is based upon work supported by the National Science Foundation Graduate Research Fellowship Program under Grant No 2444108. Any opinions, findings, and conclusions or recommendations expressed in this material are those of the author(s) and do not necessarily reflect the views of the National Science Foundation. The AlN transducer film was deposited at AMS Inc. The final fabrication steps defining the piezo actuator and DRIE were performed at the Birck Nanotechnology Center, Purdue University.

\newpage
\bibliography{main} 

\end{document}